\documentclass[prd,showkeys,onecolumn,superscriptaddress]{revtex4}
\usepackage{graphicx}
\usepackage{amsmath}

\def\gap{\;\rlap{\lower 2.5pt
 \hbox{$\sim$}}\raise 1.5pt\hbox{$>$}\;}
\def\lap{\;\rlap{\lower 2.5pt
   \hbox{$\sim$}}\raise 1.5pt\hbox{$<$}\;}
\def\gsim{\;\rlap{\lower 2.5pt
 \hbox{$\sim$}}\raise 1.5pt\hbox{$>$}\;}
\def\lsim{\;\rlap{\lower 2.5pt
   \hbox{$\sim$}}\raise 1.5pt\hbox{$<$}\;}

\def\spose#1{\hbox to 0pt{#1\hss}}
\def\lta{\mathrel{\spose{\lower 3pt\hbox{$\mathchar''218$}}
     \raise 2.0pt\hbox{$\mathchar''13C$}}}
\def\gta{\mathrel{\spose{\lower 3pt\hbox{$\mathchar''218$}}
     \raise 2.0pt\hbox{$\mathchar''13E$}}}
\newcommand{\be}{\begin{equation}}
\newcommand{\ee}{\end{equation}}

\newcommand{\ls}{\mathrel{\raise1.16pt\hbox{$<$}\kern-7.0pt 
\lower3.06pt\hbox{{$\scriptstyle \sim$}}}}         
\newcommand{\gs}{\mathrel{\raise1.16pt\hbox{$>$}\kern-7.0pt 
\lower3.06pt\hbox{{$\scriptstyle \sim$}}}}         

\long\def\comment#1{}

\def\fun#1#2{\lower3.6pt\vbox{\baselineskip0pt\lineskip.9pt
  \ialign{$\mathsurround=0pt#1\hfil##\hfil$\crcr#2\crcr\sim\crcr}}}
\def\lap{\mathrel{\mathpalette\fun <}}
\def\gap{\mathrel{\mathpalette\fun >}}
\newcommand{\ba}{\begin{eqnarray}}
\newcommand{\ea}{\end{eqnarray}}

\begin{document}
\bibliographystyle{apsrev.bst}
\title{Exact solution of Kerr-like traversable wormhole in dark matter halo}

\author{Zhaoyi Xu}
\email{zyxu@gzu.edu.cn}
\affiliation{College of Physics, Guizhou University, Guiyang 550025, China}
\affiliation{Key Laboratory of Particle Astrophysics,
Institute of High Energy Physics, Chinese Academy of Sciences,
Beijing 100049, China}

\begin{abstract}
Dark matter halos are common in galaxies and globular clusters, making it possible to use dark matter density profiles to construct transverse wormholes. 
In reference [1], an approximate solution of the traversable wormhole satisfying the isotropic pressure condition is constructed by using the dark matter density profile.
However, approximate solutions limit the discussion of traversable wormholes to a large extent.
In this work, we construct the exact solution of the traversable wormhole in the case of rotation in the dark matter halo. We have written analytical expressions for the Kerr-like traversable wormhole metric for the dark matter density profile in galaxies corresponding to different dark matter models, respectively. These rotating traversable wormholes are jointly determined by the dark matter characteristic density and the scale radius.
At the same time, we have also analyzed the energy conditions satisfied by these rotationally traversable wormholes.
\end{abstract}

\maketitle{}

\section{Introduction}
\label{intro}

Wormholes, an exact solution to Einstein's field equations in general relativity (GR), are shortcuts between different regions of spacetime in the universe, as well as passages between different universes \cite{M. Visser(1995)}. 
Since the first wormhole solution was proposed by Einstein-Rosen in 1935, wormholes have played an essential role in the study of cosmology and theories of gravity \cite{1935PhRv...48...73E}.
In the decades that followed, however, wormhole research continued to slow down. One reason may be that wormholes seem like science fiction. It was not until 1988, with the pioneering work of Thorn-Morris on wormholes, that a fresh era of wormhole research was ushered in, when Thorn-Morris presented the physical picture of transverse wormholes and their exact solutions 
\cite{1988AmJPh..56..395M,1973JMP....14..104E}. 
Since then, the research of traverse wormhole has ushered in a comprehensive and rapid development. Unfortunately, there are several problems with wormhole construction \cite{1988AmJPh..56..395M,M. Visser(1995)}. The first is that wormholes constructed in general relativity do not meet realistic energy conditions, such as null energy conditions. The second is that these wormholes, as constructed by physicists, need exotic matter, and the exotic matter needs to be massive enough to make them difficult to exist in the actual universe.
For example, some matter or energy with negative pressure is required. According to current observations and understanding of the universe, dark energy is one of the energies with negative pressure at cosmological scales, but its nature remains unclear to humans \cite{2006IJMPD..15.1753C}. These issues make the study of wormholes extremely difficult.

Here are just a few interesting examples of physicists' efforts to overcome the difficulties posed by previous research and make it possible to construct more realistic traversable wormholes \cite{2021PhRvD.103f6007M,2003PhRvD..68b4025B,2007PhRvD..76h7502R,2011PhRvL.107A1101K,2012PhRvD..85d4007K,2014PhRvD..90l6001T,2021PhRvL.126j1102B}.
Panagiota Kanti et al. constructed an ergodic wormhole in the EGB theory without exotic matter \cite{2011PhRvL.107A1101K}. Jose Luis Blazquez-Salcedo and others have constructed wormholes in EDM theories that do not require exotic matter \cite{2021PhRvL.126j1102B}.
On the other hand, given the large mass of dark matter halos in galaxies, physicists have investigated the possibility of a wormhole at the center of galaxies. This fact led to a series of works. 
For example, Farook Rahaman et al used the NFW dark matter distribution of the cold dark matter model to construct wormhole solutions and analyze their properties \cite{2014EPJC...74.2750R}. Kimet Jusufi et al. discuss the possibility of wormhole formation in the TF dark-matter halo of FDM \cite{2019GReGr..51..102J}.
Zhaoyi Xu et al. discussed the possibility of spherical and axisymmetric wormholes forming in the dark matter distribution under isotropic pressure \cite{2020EPJC...80...70X}. However, the wormhole solutions constructed by Zhaoyi Xu et al. are relatively limited in the axisymmetric case, because they assume some special conditions $(f=g)$.
By inspection, it is found that this condition can only hold in a truly narrow parameter space, and hence the solution of the axisymmetric wormhole requires additional in-depth analysis and discussion. 
This work attempts to address the more general problem of axisymmetric wormhole structures in dark matter halos when isotropic pressure conditions are satisfied.

In this work, we construct exact solutions for the axisymmetric case of a transverse wormhole in a dark matter halo under various isotropic conditions and discuss their energy conditions tentatively.
In section \ref{intro}, we introduce the space-time metric and properties of non-rotationally traversable wormholes. In the \ref{rotat} section, the exact solution of the rotation wormhole is derived under the isotropic condition. In section \ref{properties}, the energy conditions of these wormholes are analyzed. The \ref{discuss} section is the summary.

\section{Non-rotating transversible wormhole}
\label{intro}

For simplicity, we consider a static non-rotating traversable wormhole spacetime metric, where the spacetime geometrical line elements should contain no time coordinates, that is, the metric coefficients are only functions of the spatial coordinates.
In general, the space-time line element corresponding to a non-rotating traversable wormhole should be
\begin{equation}
ds^{2}=-e^{2\Phi}dt^{2}+\dfrac{1}{1-\frac{b(r)}{r}}dr^{2}+h(r)(d\theta^{2}+\sin^{2}\theta d\phi^{2}),
\label{SPTW1}
\end{equation}
where $\Phi(r)$ is a redshift function, which is related to the gravitational redshift of the traversable wormhole spacetime, and $\Phi(r)$ can only be finite value; 
$b(r)$ is the shape function, which is associated with the geometry of the traversable wormhole (which can be represented by the embedding diagram of the traversable wormhole in three-dimensional space);
$h(r)$ is a spherical shape function that determines whether an traversable wormhole of a certain radius at a certain time has spherical geometric properties.
According to the geometry of traversable wormholes, the radial coordinate $r$ takes values from the throat of the traversable wormhole to the flat spacetime at infinity.

The space-time element (\ref{SPTW1}) is universally satisfied, and all non-rotating traversable wormholes can be written in this geometry.
In order to obtain the analytical expression of unknown functions $\Phi(r)$, $b(r)$ and $h(r)$ in space-time line element (\ref{SPTW1}) under specific circumstances. We qualify again the space-time line element (\ref{SPTW1}). If the embedded diagram is spherical in three-dimensional space through traversable wormhole, then $h(r)=r^2$.
Space-time line elements need to satisfy Einstein's field equations $G_{\mu\nu}=8\pi T_{\mu\nu}$. For common ideal fluids, the energy-momentum tensor can be reduced to the following diagonal form $T_{\mu\nu}=diag(-\rho,P_{r},P_{\theta},P_{\phi})$. 
By substituting the non-rotating space-time linear element (\ref{SPTW1}) that can pass through the traversable wormhole into the gravitational field equation, Einstein's field equation can be converted into the following equations
\begin{equation}
8\pi\rho(r)=\dfrac{b^{'}(r)}{r^{2}},
\label{SPTW2}
\end{equation}
\begin{equation}
8\pi P_{r}(r)=-\dfrac{b(r)}{r^{3}}+2\Big(1-\dfrac{b(r)}{r}\Big)\dfrac{\Phi^{'}(r)}{r},
\label{SPTW3}
\end{equation}
\begin{equation}
8\pi P_{\theta}(r)=8\pi P_{\phi}(r)=\Big(1-\dfrac{b(r)}{r}\Big)\Big[\Phi^{''}(r)+\Phi^{'2}(r)-\dfrac{rb^{'}(r)-b(r)}{2r(r-b(r))}\Phi^{'}(r)-\dfrac{rb^{'}(r)-b(r)}{2r^{2}(r-b(r))}+\dfrac{\Phi^{'}(r)}{r}\Big].
\label{SPTW4}
\end{equation}

Given the density profile of dark matter, the spherically symmetric traversable wormhole metric can be obtained if the isothermal condition $P_{r}=P_{\theta}=P_{\phi}=P$ is taken into account.
In literature \cite{2020EPJC...80...70X}, the author has given three common traversable wormhole space-time line elements under the density profile of dark matter, and the analytical expression are
\begin{equation}
e^{2\Phi(r)}=\left\{
             \begin{aligned}
            & \Big(\dfrac{r+R_{s}}{r_{0}+R_{s}}\Big)^{\dfrac{4\omega}{1+\omega}}\times\Big(\dfrac{r}{r_{0}}\Big)^{\dfrac{2\omega}{1+\omega}},  \: \: \: \: \: {\rm for \, NFW \, profile;} & & \\
            & \Big(\dfrac{r}{r_{0}}\dfrac{\sin(\dfrac{\pi r_{0}}{R})}{\sin(\dfrac{\pi r}{R})}\Big)^{\dfrac{2\omega}{1+\omega}}, \: \: \: \: \: {\rm for\, TF \, profile;} & & \\
            & \Big(\dfrac{r^{2}+R^{2}_{c}}{r^{2}_{0}+R^{2}_{c}}\Big)^{\dfrac{2\omega}{1+\omega}}, \: \: \: \: \: {\rm for\, PI \, profile.} & &
             \end{aligned}
\right.
\label{SPTW5}
\end{equation}

\begin{equation}
b(r)=\left\{
             \begin{aligned}
           &  r_{0}+8\pi\rho_{s}R^{3}_{s}[\ln\dfrac{r+R_{s}}{r_{0}+R_{s}}+R_{s}(\dfrac{1}{r+R_{s}}-\dfrac{1}{r_{0}+R_{s}})]   ,  \: \: \: \: \: {\rm for \, NFW \, profile;} & & \\
           &  r_{0}+\dfrac{8\rho_{0}R^{3}}{\pi^{2}}[\sin(\dfrac{\pi r}{R})-\sin(\dfrac{\pi r_{0}}{R})]-\dfrac{8\rho_{0}R^{3}}{\pi}[\dfrac{r}{R}\cos(\dfrac{\pi r}{R})-\dfrac{r_{0}}{R}\cos(\dfrac{\pi r_{0}}{R})], \: \: \: \: \: {\rm for\, TF \, profile;} & & \\
           &  r_{0}+8\pi\rho_{0}R^{2}_{c}[r-r_{0}-R^{2}_{c}\ln\dfrac{r+R^{2}_{c}}{r_{0}+R^{2}_{c}}]  , \: \: \: \: \: {\rm for\, PI \, profile.} & &
             \end{aligned}
\right.
\label{SPTW6}
\end{equation}
Where $\omega$ is the equation of state of dark matter, and it is the ratio of pressure $P$ to density $\rho$ of dark matter. The physical significance of other model parameters is related to the dark matter model \cite{2020EPJC...80...70X}.

In reference \cite{2020EPJC...80...70X}, they generalize the spherically symmetric traversable wormhole solution to the rotating case based on the condition $e^{2\Phi(r)}\approx 1-b(r)/r$.
However, if the condition $e^{2\Phi(r)}\approx 1-b(r)/r$ is satisfied at any radial radius, then the parameters of the dark matter model and the equation of state cannot be chosen arbitrarily, which considerably limits the physical connotation of the rotationally traversable wormhole solution described in Ref.\cite{2020EPJC...80...70X}.
In the following calculations, we will discard this constraint and derive a more general rotationally traversable wormhole solution.

\section{Rotating transversible wormhole}
\label{rotat}
Before deriving the geometrical line elements of rotationally traversable wormholes, we need to recall the preconditions of this work, namely, the isotropic condition on the dark matter pressure.
The isotropic condition for the dark matter pressure states that at any point in the manifold of wormhole spacetime, the pressure generated by the dark matter is equal in different coordinate directions.
For the wormhole (\ref{SPTW5} and \ref{SPTW6}), the matter pressure corresponding to it is
\begin{equation}
P=P_{r}=P_{\theta}=P_{\phi}=\omega\rho=\left\{
             \begin{aligned}
           &  \dfrac{\omega\rho_{s}}{\dfrac{r}{R_{s}}(1+\dfrac{r}{R_{s}})^{2}}   ,  \: \: \: \: \: {\rm for \, NFW \, profile;} & & \\
           &  \omega\rho_{s}\dfrac{\sin(kr)}{kr}, \: \: \: \: \: {\rm for\, TF \, profile;} & & \\
           &  \dfrac{\omega\rho_{0}}{1+(\dfrac{r}{R_{c}})^{2}}  , \: \: \: \: \: {\rm for\, PI \, profile.} & &
             \end{aligned}
\right.
\label{SPTW6}
\end{equation}
When spherically symmetric traversable wormhole space-time satisfies this condition, Newman-Janis algorithm can be used to solve the rotating traversable wormhole solution.
For an introduction to the Newman-Janis algorithm, a detailed discussion can be found in the relevant literature \cite{1965JMP.....6..915N,2014EPJC...74.2865A,2014PhLB..730...95A,2016EPJC...76....3A,2016EPJC...76....7A}.   
The main idea of this algorithm is to extend the spherically symmetric space-time metric to the axisymmetric case by using complex coordinate transformation, and transform Einstein's field equation into a second order partial differential equations. By solving the partial differential equations, the analytic expression of the unknown function in the axisymmetric space-time metric can be obtained.
Therefore, for a given spherically symmetric wormhole metric, as long as its metric meets the solvable condition (the corresponding second-order partial differential equations are solvable), the analytical expression of the corresponding axisymmetric wormhole metric can be given.
Calculations show that the traversable wormhole (\ref{SPTW5} and \ref{SPTW6}) satisfies the solvability condition only. Next, the rotationally traversable wormhole metric is derived following the standard procedure of the NJ algorithm.

By a series of calculations, it was found that the general form of the axisymmetric wormhole corresponding to the traversable wormhole (\ref{SPTW5} and \ref{SPTW6}) is
\begin{equation}
ds^{2}=-\dfrac{\Psi}{\Sigma^{2}}\big(1-\dfrac{2\overline{f}}{\Sigma^{2}} \big)dt^{2}+\dfrac{\Psi}{\Delta}dr^{2}-\dfrac{4a\overline{f}\sin^{2}\theta \Psi}{\Sigma^{4}}dtd\phi+\Psi d\theta^{2}+\dfrac{\Psi A \sin^{2}\theta}{\Sigma^{4}}d\phi^{2},
\label{NJAWH1}
\end{equation}
where 
\begin{equation}
k(r)=r^{2}\sqrt{\frac{e^{2\Phi}}{1-\frac{b(r)}{r}}},
\label{NJAWH2}
\end{equation}
\begin{equation}
2\overline{f}=r^{2}\sqrt{\frac{e^{2\Phi}}{1-\frac{b(r)}{r}}}-r^{2}e^{2\Phi},
\label{NJAWH3}
\end{equation}
\begin{equation}
\Delta(r)=r^{2}e^{2\Phi}+a^{2},
\label{NJAWH4}
\end{equation}
\begin{equation}
A=\Big(r^{2}\sqrt{\frac{e^{2\Phi}}{1-\frac{b(r)}{r}}}+a^{2}\Big)^{2}-a^{2}\Delta \sin^{2}\theta,
\label{NJAWH5}
\end{equation}
\begin{equation}
\Sigma^{2}=r^{2}\sqrt{\frac{e^{2\Phi}}{1-\frac{b(r)}{r}}}+a^{2}\cos^{2}\theta,
\label{NJAWH6}
\end{equation}
and $\Psi=\Psi(r,\theta,a)$ is an unknown function in a rotating wormhole. 

So far we have not solved Einstein's field equations. 
If the rotating wormhole (\ref{NJAWH1}) is a solution of the Einstein field equations, then the rotating wormhole (\ref{NJAWH1}) can be used to simplify the Einstein field equations and the axisymmetry condition, thus we can reduce the Einstein field equations to ($y=\cos\theta$)
\begin{equation}
(k+a^{2}y^{2})^{2}(3\frac{\partial \Psi}{\partial r}\frac{\partial \Psi}{\partial y^{2}}-2\Psi\frac{\partial^{2}\Psi}{\partial r \partial y^{2}})=3a^{2}k_{,r}\Psi^{2},
\label{NJAWH7}
\end{equation}
\begin{equation}
\Psi[k^{2}_{,r}+k(2-k_{,rr})-a^{2}y^{2}(2+k_{,rr})]+(k+a^{2}y^{2})(4y^{2}\frac{\partial \Psi}{\partial y^{2}}-k_{,r}\frac{\partial \Psi}{\partial r})=0.
\label{NJAWH8}
\end{equation}
This system of equations is extremely complex and difficult to solve under normal circumstances. However, when the isotropic condition is satisfied, the analytical form of $\Psi$ can be obtained. The analytical form of $\Psi$ is closely related to the pressure of spherically symmetric wormhole.
For reference \cite{2014PhLB..730...95A}, the solution of the equations (\ref{NJAWH7} and \ref{NJAWH8}) can be obtained, and its expression is
\begin{equation}
\Psi=r^{2}+P^{2}+a^{2}\cos^{2}\theta.
\label{NJA15}
\end{equation}
So far we have obtained rotationally traversable wormhole solutions corresponding to the dark matter density profile, and we will write down their specific expressions for the convenience of the following analysis and discussion.

I. NFW density profile and CDM halo
\begin{equation}
\Psi_{\rm{NFW}}=r^{2}+a^{2}\cos^{2}\theta+\Big(\dfrac{\omega\rho_{s}}{\dfrac{r}{R_{s}}(1+\dfrac{r}{R_{s}})^{2}}\Big)^{2},
\label{NFW-TW1}
\end{equation}
\begin{equation}
k_{\rm{NFW}}=r^{2}\sqrt{\frac{\Big(\dfrac{r+R_{s}}{r_{0}+R_{s}}\Big)^{\dfrac{4\omega}{1+\omega}}\times\Big(\dfrac{r}{r_{0}}\Big)^{\dfrac{2\omega}{1+\omega}}}{1-\frac{r_{0}+8\pi\rho_{s}R^{3}_{s}\Big[\ln\dfrac{r+R_{s}}{r_{0}+R_{s}}+R_{s}(\dfrac{1}{r+R_{s}}-\dfrac{1}{r_{0}+R_{s}})\Big]}{r}}},
\label{NFW-TW2}
\end{equation}
\begin{equation}
\Sigma^{2}_{\rm{NFW}}=r^{2}\sqrt{\frac{\Big(\dfrac{r+R_{s}}{r_{0}+R_{s}}\Big)^{\dfrac{4\omega}{1+\omega}}\times\Big(\dfrac{r}{r_{0}}\Big)^{\dfrac{2\omega}{1+\omega}}}{1-\frac{r_{0}+8\pi\rho_{s}R^{3}_{s}\Big[\ln\dfrac{r+R_{s}}{r_{0}+R_{s}}+R_{s}(\dfrac{1}{r+R_{s}}-\dfrac{1}{r_{0}+R_{s}})\Big]}{r}}}+a^{2}\cos^{2}\theta,
\label{NFW-TW3}
\end{equation}
\begin{equation}
2\overline{f}_{\rm{NFW}}=r^{2}\sqrt{\frac{\Big(\dfrac{r+R_{s}}{r_{0}+R_{s}}\Big)^{\dfrac{4\omega}{1+\omega}}\times\Big(\dfrac{r}{r_{0}}\Big)^{\dfrac{2\omega}{1+\omega}}}{1-\frac{r_{0}+8\pi\rho_{s}R^{3}_{s}\Big[\ln\dfrac{r+R_{s}}{r_{0}+R_{s}}+R_{s}(\dfrac{1}{r+R_{s}}-\dfrac{1}{r_{0}+R_{s}})\Big]}{r}}}-r^{2}\Big(\dfrac{r+R_{s}}{r_{0}+R_{s}}\Big)^{\dfrac{4\omega}{1+\omega}}\times\Big(\dfrac{r}{r_{0}}\Big)^{\dfrac{2\omega}{1+\omega}},
\label{NFW-TW4}
\end{equation}
\begin{equation}
\Delta_{\rm{NFW}}=r^{2}\Big(\dfrac{r+R_{s}}{r_{0}+R_{s}}\Big)^{\dfrac{4\omega}{1+\omega}}\times\Big(\dfrac{r}{r_{0}}\Big)^{\dfrac{2\omega}{1+\omega}}+a^{2},
\label{NFW-TW5}
\end{equation}
\begin{equation}
A_{\rm{NFW}}=\Big(r^{2}\sqrt{\frac{\Big(\dfrac{r+R_{s}}{r_{0}+R_{s}}\Big)^{\dfrac{4\omega}{1+\omega}}\times\Big(\dfrac{r}{r_{0}}\Big)^{\dfrac{2\omega}{1+\omega}}}{1-\frac{r_{0}+8\pi\rho_{s}R^{3}_{s}\Big[\ln\dfrac{r+R_{s}}{r_{0}+R_{s}}+R_{s}(\dfrac{1}{r+R_{s}}-\dfrac{1}{r_{0}+R_{s}})\Big]}{r}}}+a^{2}\Big)^{2}-a^{2}\sin^{2}\theta\Big(r^{2}\Big(\dfrac{r+R_{s}}{r_{0}+R_{s}}\Big)^{\dfrac{4\omega}{1+\omega}}\times\Big(\dfrac{r}{r_{0}}\Big)^{\dfrac{2\omega}{1+\omega}}+a^{2}\Big),
\label{NFW-TW6}
\end{equation}

II. TF density profile and FDM halo
\begin{equation}
\Psi_{\rm{TF}}=r^{2}+a^{2}\cos^{2}\theta+\Big(\omega\rho_{s}\dfrac{\sin(kr)}{kr}\Big)^{2},
\label{TF-TW1}
\end{equation}
\begin{equation}
k_{\rm{TF}}=r^{2}\sqrt{\frac{\Big(\dfrac{r}{r_{0}}\dfrac{\sin(\dfrac{\pi r_{0}}{R})}{\sin(\dfrac{\pi r}{R})}\Big)^{\dfrac{2\omega}{1+\omega}}}{1-\frac{r_{0}+\dfrac{8\rho_{0}R^{3}}{\pi^{2}}[\sin(\dfrac{\pi r}{R})-\sin(\dfrac{\pi r_{0}}{R})]-\dfrac{8\rho_{0}R^{3}}{\pi}[\dfrac{r}{R}\cos(\dfrac{\pi r}{R})-\dfrac{r_{0}}{R}\cos(\dfrac{\pi r_{0}}{R})]}{r}}},
\label{TF-TW2}
\end{equation}
\begin{equation}
\Sigma^{2}_{\rm{TF}}=r^{2}\sqrt{\frac{\Big(\dfrac{r}{r_{0}}\dfrac{\sin(\dfrac{\pi r_{0}}{R})}{\sin(\dfrac{\pi r}{R})}\Big)^{\dfrac{2\omega}{1+\omega}}}{1-\frac{r_{0}+\dfrac{8\rho_{0}R^{3}}{\pi^{2}}[\sin(\dfrac{\pi r}{R})-\sin(\dfrac{\pi r_{0}}{R})]-\dfrac{8\rho_{0}R^{3}}{\pi}[\dfrac{r}{R}\cos(\dfrac{\pi r}{R})-\dfrac{r_{0}}{R}\cos(\dfrac{\pi r_{0}}{R})]}{r}}}+a^{2}\cos^{2}\theta,
\label{TF-TW3}
\end{equation}
\begin{equation}
2\overline{f}_{\rm{TF}}=r^{2}\sqrt{\frac{\Big(\dfrac{r}{r_{0}}\dfrac{\sin(\dfrac{\pi r_{0}}{R})}{\sin(\dfrac{\pi r}{R})}\Big)^{\dfrac{2\omega}{1+\omega}}}{1-\frac{r_{0}+\dfrac{8\rho_{0}R^{3}}{\pi^{2}}[\sin(\dfrac{\pi r}{R})-\sin(\dfrac{\pi r_{0}}{R})]-\dfrac{8\rho_{0}R^{3}}{\pi}[\dfrac{r}{R}\cos(\dfrac{\pi r}{R})-\dfrac{r_{0}}{R}\cos(\dfrac{\pi r_{0}}{R})]}{r}}}-r^{2}\Big(\dfrac{r}{r_{0}}\dfrac{\sin(\dfrac{\pi r_{0}}{R})}{\sin(\dfrac{\pi r}{R})}\Big)^{\dfrac{2\omega}{1+\omega}},
\label{TF-TW4}
\end{equation}
\begin{equation}
\Delta_{\rm{TF}}=r^{2}\Big(\dfrac{r}{r_{0}}\dfrac{\sin(\dfrac{\pi r_{0}}{R})}{\sin(\dfrac{\pi r}{R})}\Big)^{\dfrac{2\omega}{1+\omega}}+a^{2},
\label{TF-TW5}
\end{equation}
\begin{equation}
A_{\rm{TF}}=\Big(r^{2}\sqrt{\frac{\Big(\dfrac{r}{r_{0}}\dfrac{\sin(\dfrac{\pi r_{0}}{R})}{\sin(\dfrac{\pi r}{R})}\Big)^{\dfrac{2\omega}{1+\omega}}}{1-\frac{r_{0}+\dfrac{8\rho_{0}R^{3}}{\pi^{2}}[\sin(\dfrac{\pi r}{R})-\sin(\dfrac{\pi r_{0}}{R})]-\dfrac{8\rho_{0}R^{3}}{\pi}[\dfrac{r}{R}\cos(\dfrac{\pi r}{R})-\dfrac{r_{0}}{R}\cos(\dfrac{\pi r_{0}}{R})]}{r}}}+a^{2}\Big)^{2}-a^{2}\sin^{2}\theta\Big(r^{2}\Big(\dfrac{r}{r_{0}}\dfrac{\sin(\dfrac{\pi r_{0}}{R})}{\sin(\dfrac{\pi r}{R})}\Big)^{\dfrac{2\omega}{1+\omega}}+a^{2}\Big),
\label{TF-TW6}
\end{equation}

III. PI density profile and MOND halo
\begin{equation}
\Psi_{\rm{PI}}=r^{2}+a^{2}\cos^{2}\theta+\Big(\dfrac{\omega\rho_{0}}{1+(\dfrac{r}{R_{c}})^{2}}\Big)^{2},
\label{PI-TW1}
\end{equation}
\begin{equation}
k_{\rm{PI}}=r^{2}\sqrt{\frac{\Big(\dfrac{r^{2}+R^{2}_{c}}{r^{2}_{0}+R^{2}_{c}}\Big)^{\dfrac{2\omega}{1+\omega}}}{1-\frac{r_{0}+8\pi\rho_{0}R^{2}_{c}[r-r_{0}-R^{2}_{c}\ln\dfrac{r+R^{2}_{c}}{r_{0}+R^{2}_{c}}]}{r}}},
\label{PI-TW2}
\end{equation}
\begin{equation}
\Sigma^{2}_{\rm{PI}}=r^{2}\sqrt{\frac{\Big(\dfrac{r^{2}+R^{2}_{c}}{r^{2}_{0}+R^{2}_{c}}\Big)^{\dfrac{2\omega}{1+\omega}}}{1-\frac{r_{0}+8\pi\rho_{0}R^{2}_{c}[r-r_{0}-R^{2}_{c}\ln\dfrac{r+R^{2}_{c}}{r_{0}+R^{2}_{c}}]}{r}}}+a^{2}\cos^{2}\theta,
\label{PI-TW3}
\end{equation}
\begin{equation}
2\overline{f}_{\rm{PI}}=r^{2}\sqrt{\frac{\Big(\dfrac{r^{2}+R^{2}_{c}}{r^{2}_{0}+R^{2}_{c}}\Big)^{\dfrac{2\omega}{1+\omega}}}{1-\frac{r_{0}+8\pi\rho_{0}R^{2}_{c}[r-r_{0}-R^{2}_{c}\ln\dfrac{r+R^{2}_{c}}{r_{0}+R^{2}_{c}}]}{r}}}-r^{2}\Big(\dfrac{r^{2}+R^{2}_{c}}{r^{2}_{0}+R^{2}_{c}}\Big)^{\dfrac{2\omega}{1+\omega}},
\label{PI-TW4}
\end{equation}
\begin{equation}
\Delta_{\rm{PI}}=r^{2}\Big(\dfrac{r^{2}+R^{2}_{c}}{r^{2}_{0}+R^{2}_{c}}\Big)^{\dfrac{2\omega}{1+\omega}}+a^{2},
\label{PI-TW5}
\end{equation}
\begin{equation}
A_{\rm{PI}}=\Big(r^{2}\sqrt{\frac{\Big(\dfrac{r^{2}+R^{2}_{c}}{r^{2}_{0}+R^{2}_{c}}\Big)^{\dfrac{2\omega}{1+\omega}}}{1-\frac{r_{0}+8\pi\rho_{0}R^{2}_{c}[r-r_{0}-R^{2}_{c}\ln\dfrac{r+R^{2}_{c}}{r_{0}+R^{2}_{c}}]}{r}}}+a^{2}\Big)^{2}-a^{2}\sin^{2}\theta\Big(r^{2}\Big(\dfrac{r^{2}+R^{2}_{c}}{r^{2}_{0}+R^{2}_{c}}\Big)^{\dfrac{2\omega}{1+\omega}}+a^{2}\Big),
\label{PI-TW6}
\end{equation}

\section{Energy conditions}
\label{properties}

This is extremely crucial for the construction of traversable wormholes satisfying various energy conditions, since an traversable wormholes can only be real if it satisfies certain energy conditions.
The energy condition actually describes the properties of the energy density and pressure in the spacetime, in other words it describes the properties of the energy-momentum tensor distribution in the spacetime.
If the energy condition is to be discussed, the non-zero energy-momentum tensor component of the solution for a rotating ergodic wormhole needs to be computed. The non-zero components of the energy-momentum tensor are as follows \cite{2014PhLB..730...95A}
\begin{equation}
\rho_{\rm{Kerr-like WH}}=\frac{1}{\Sigma^{6}}\Big[-6e^{2\Phi(r)}+r^{2}+p^{2}+a^{2}(2-\cos^{2}\theta)\Big]p^{2}-\frac{2}{\Sigma^{4}}\Big(r(e^{2\Phi(r)})^{'} -e^{2\Phi(r)}\Big),
\label{EC1}
\end{equation}
\begin{equation}
p_{r\rm{Kerr-like WH}}=\frac{1}{\Sigma^{6}}\Big[2e^{2\Phi(r)}+r^{2}+p^{2}+a^{2}\cos^{2}\theta\Big]p^{2}-\frac{2}{\Sigma^{4}}\Big(r(e^{2\Phi(r)})^{'} -e^{2\Phi(r)}\Big),
\label{EC2}
\end{equation}
\begin{equation}
p_{\theta\rm{Kerr-like WH}}=\frac{2(r^{2}+a^{2}\cos^{2}\theta)}{\Sigma^{6}}e^{2\Phi(r)}-\frac{1}{\Sigma^{4}}\Big[2r(e^{2\Phi(r)})^{'}+p^{2}\Big]+\frac{1}{\Sigma^{2}}(e^{2\Phi(r)})^{''} ,
\label{EC3}
\end{equation}
\begin{equation}
p_{\phi\rm{Kerr-like WH}}=\frac{2}{\Sigma^{6}}[(r^{2}+a^{2}\cos^{2}\theta)e^{2\Phi(r)}-2a^{2}\sin^{2}\theta p^{2}]--\frac{1}{\Sigma^{4}}\Big[2r(e^{2\Phi(r)})^{'}+p^{2}\Big]+\frac{1}{\Sigma^{2}}(e^{2\Phi(r)})^{''}.
\label{EC4}
\end{equation}

From the analysis of Eq. (\ref{EC1}), it can be found that the energy density is positive for all three types of rotating wormholes, which is somewhat natural since for spherically symmetric wormholes we are talking in terms of positive dark matter density. For wormholes, we are mostly concerned with the null energy condition (NEC).
\begin{equation}
\rho_{\rm{WH}}+p_{r\rm{WH}}=\left\{
\begin{aligned}
       &  \frac{2P^{2}}{\Sigma^{6}}\Big(2\Big(\dfrac{r+R_{s}}{r_{0}+R_{s}}\Big)^{\dfrac{4\omega}{1+\omega}}\times\Big(\dfrac{r}{r_{0}}\Big)^{\dfrac{2\omega}{1+\omega}}-r^{2}-\Big[\dfrac{\omega\rho_{s}}{\dfrac{r}{R_{s}}(1+\dfrac{r}{R_{s}})^{2}}\Big]^{2}-a^{2}\Big),  \: \: \: \: \: {\rm for \, NFW \, profile;} & & \\
          &   \frac{2P^{2}}{\Sigma^{6}}\Big(2\Big(\dfrac{r}{r_{0}}\dfrac{\sin(\dfrac{\pi r_{0}}{R})}{\sin(\dfrac{\pi r}{R})}\Big)^{\dfrac{2\omega}{1+\omega}}-r^{2}-\Big[\omega\rho_{s}\dfrac{\sin(kr)}{kr}\Big]^{2}-a^{2}\Big), \: \: \: \: \: {\rm for\, TF \, profile;} & & \\
           &  \frac{2P^{2}}{\Sigma^{6}}\Big(2\Big(\dfrac{r^{2}+R^{2}_{c}}{r^{2}_{0}+R^{2}_{c}}\Big)^{\dfrac{2\omega}{1+\omega}}-r^{2}-\Big[\dfrac{\omega\rho_{0}}{1+(\dfrac{r}{R_{c}})^{2}}\Big]^{2}-a^{2}\Big), \: \: \: \: \: {\rm for\, PI \, profile.} & &
             \end{aligned}
\right.
\label{EC5}
\end{equation}
By numerically calculating the null energy condition (\ref{EC5}), we find the following conclusion (In addition, the condition $\rho_{\rm{WH}}+p_{\theta\rm{WH}}$ and $\rho_{\rm{WH}}+p_{\phi\rm{WH}}$ can be added). 
When $r$ approaches zero, the wormhole corresponding to NFW does not satisfy NEC, but the wormhole corresponding to TF and PI does. 
As $r$ approaches infinity, none of the three types of wormholes satisfy NEC. 
For TF and PI wormholes, the wormhole space-time only satisfies NEC in a certain region, and the critical radius $r_{c}$ of these wormholes satisfying NEC and not satisfying NEC is determined by the following equation.
For the TF density profile and FDM halo, the critical radius satisfies the equation $2\Big(\dfrac{r_{c}}{r_{0}}\dfrac{\sin(\dfrac{\pi r_{0}}{R})}{\sin(\dfrac{\pi r_{c}}{R})}\Big)^{\dfrac{2\omega}{1+\omega}}-r_{c}^{2}-\Big[\omega\rho_{s}\dfrac{\sin(kr_{c})}{kr_{c}}\Big]^{2}-a^{2}=0$. For the PI density profile and MOND halo, the critical radius satisfies the equation $2\Big(\dfrac{r_{c}^{2}+R^{2}_{c}}{r^{2}_{0}+R^{2}_{c}}\Big)^{\dfrac{2\omega}{1+\omega}}-r_{c}^{2}-\Big[\dfrac{\omega\rho_{0}}{1+(\dfrac{r_{c}}{R_{c}})^{2}}\Big]^{2}-a^{2}=0$.

\section{Summary}
\label{discuss}

In this work, we construct a rotating traversable wormhole solution in a dark matter halo when the pressure profile of the dark matter halo satisfies the isotropic condition. 
These rotating traversable wormhole solutions are generalizations of the wormhole solutions in literature \cite{2020EPJC...80...70X}, and can satisfy further space-time regions and parameter ranges.
However, in the general case of rotating traversable wormhole spacetimes in dark matter halos, the distribution of dark matter supporting traversable wormholes will no longer satisfy the isotropy condition, which is precisely the case for dark matter halos in realistic environments.
Through calculation, we find that the solutions of these rotating traversable wormholes meet the conditions of null energy condition and weak energy condition in a certain space-time region, which makes these traversable wormhole solutions have more realistic physical significance.
Since the space-time structure of these rotating traversable wormholes is particularly complex, the properties of these rotating traversable wormholes will be discussed in the following work. Our purpose is to use dark matter and other more realistic physical conditions to construct traversable wormhole solutions that satisfy the energy condition and do not require exotic matter, and thus shed new light on the study of traversable wormholes.

\begin{acknowledgments}
We acknowledge the anonymous referee for a constructive report that has significantly improved this paper. We acknowledge the  Special Natural Science Fund of Guizhou University (grant
No. X2020068) and the financial support from the China Postdoctoral Science Foundation funded project under grants No. 2019M650846.
\end{acknowledgments}


\begin{thebibliography}{99}

\bibitem[Xu et al.(2020)]{2020EPJC...80...70X} Xu, Z., Tang, M., Cao, G., Zhang, S.-N.\ 2020.\ Possibility of traversable wormhole formation in the dark matter halo with istropic pressure.\ European Physical Journal C 80. doi:10.1140/epjc/s10052-020-7636-0

\bibitem[M. Visser(1995)]{M. Visser(1995)}  M. Visser
.\ 1995.\  Lorentzian Wormholes: From Einstein to Hawking.\  (AIP, Woodbury, NY, 1995), p. 412.

\bibitem[Einstein and Rosen(1935)]{1935PhRv...48...73E} Einstein, A., Rosen, N.\ 1935.\ The Particle Problem in the General Theory of Relativity.\ Physical Review 48, 73–77. doi:10.1103/PhysRev.48.73

\bibitem[Morris and Thorne(1988)]{1988AmJPh..56..395M} Morris, M.~S., Thorne, K.~S.\ 1988.\ Wormholes in spacetime and their use for interstellar travel: A tool for teaching general relativity.\ American Journal of Physics 56, 395–412. doi:10.1119/1.15620

\bibitem[Ellis(1973)]{1973JMP....14..104E} Ellis, H.~G.\ 1973.\ Ether flow through a drainhole: A particle model in general relativity.\ Journal of Mathematical Physics 14, 104–118. doi:10.1063/1.1666161

\bibitem[Copeland et al.(2006)]{2006IJMPD..15.1753C} Copeland, E.~J., Sami, M., Tsujikawa, S.\ 2006.\ Dynamics of Dark Energy.\ International Journal of Modern Physics D 15, 1753–1935. doi:10.1142/S021827180600942X


\bibitem[Maldacena and Milekhin(2021)]{2021PhRvD.103f6007M} Maldacena, J., Milekhin, A.\ 2021.\ Humanly traversable wormholes.\ Physical Review D 103. doi:10.1103/PhysRevD.103.066007

\bibitem[Bronnikov et al.(2003)]{2003PhRvD..68b4025B} Bronnikov, K., Melnikov, V., Dehnen, H.\ 2003.\ General class of brane-world black holes.\ Physical Review D 68. doi:10.1103/PhysRevD.68.024025

\bibitem[Richarte and Simeone(2007)]{2007PhRvD..76h7502R} Richarte, M.~G., Simeone, C.\ 2007.\ Thin-shell wormholes supported by ordinary matter in Einstein-Gauss-Bonnet gravity.\ Physical Review D 76. doi:10.1103/PhysRevD.76.087502

\bibitem[Kanti et al.(2011)]{2011PhRvL.107A1101K} Kanti, P., Kleihaus, B., Kunz, J.\ 2011.\ Wormholes in Dilatonic Einstein-Gauss-Bonnet Theory.\ Physical Review Letters 107. doi:10.1103/PhysRevLett.107.271101

\bibitem[Kanti et al.(2012)]{2012PhRvD..85d4007K} Kanti, P., Kleihaus, B., Kunz, J.\ 2012.\ Stable Lorentzian wormholes in dilatonic Einstein-Gauss-Bonnet theory.\ Physical Review D 85. doi:10.1103/PhysRevD.85.044007

\bibitem[Tomikawa et al.(2014)]{2014PhRvD..90l6001T} Tomikawa, Y., Shiromizu, T., Izumi, K.\ 2014.\ Wormhole on DGP brane.\ Physical Review D 90. doi:10.1103/PhysRevD.90.126001

\bibitem[Bl{\'a}zquez-Salcedo et al.(2021)]{2021PhRvL.126j1102B} Bl{\'a}zquez-Salcedo, J.~L., Knoll, C., Radu, E.\ 2021.\ Traversable Wormholes in Einstein-Dirac-Maxwell Theory.\ Physical Review Letters 126. doi:10.1103/PhysRevLett.126.101102

\bibitem[Rahaman et al.(2014)]{2014EPJC...74.2750R} Rahaman, F., Kuhfittig, P.~K.~F., Ray, S., Islam, N.\ 2014.\ Possible existence of wormholes in the galactic halo region.\ European Physical Journal C 74. doi:10.1140/epjc/s10052-014-2750-5

\bibitem[Jusufi et al.(2019)]{2019GReGr..51..102J} Jusufi, K., Jamil, M., Rizwan, M.\ 2019.\ On the possibility of wormhole formation in the galactic halo due to dark matter Bose-Einstein condensates.\ General Relativity and Gravitation 51. doi:10.1007/s10714-019-2586-2

\bibitem[Newman and Janis(1965)]{1965JMP.....6..915N} Newman, E.~T., Janis, A.~I.\ 1965.\ Note on the Kerr Spinning-Particle Metric.\ Journal of Mathematical Physics 6, 915–917. doi:10.1063/1.1704350

\bibitem[Azreg-A{\"\i}nou(2014)]{2014EPJC...74.2865A} Azreg-A{\"\i}nou, M.\ 2014.\ From static to rotating to conformal static solutions: rotating imperfect fluid wormholes with(out) electric or magnetic field.\ European Physical Journal C 74. doi:10.1140/epjc/s10052-014-2865-8

\bibitem[Azreg-A{\"\i}nou(2014)]{2014PhLB..730...95A} Azreg-A{\"\i}nou, M.\ 2014.\ Regular and conformal regular cores for static and rotating solutions.\ Physics Letters B 730, 95–98. doi:10.1016/j.physletb.2014.01.041

\bibitem[Azreg-A{\"\i}nou(2016)]{2016EPJC...76....3A} Azreg-A{\"\i}nou, M.\ 2016.\ Wormhole solutions sourced by fluids, I: Two-fluid charged sources.\ European Physical Journal C 76. doi:10.1140/epjc/s10052-015-3835-5

\bibitem[Azreg-A{\"\i}nou(2016)]{2016EPJC...76....7A} Azreg-A{\"\i}nou, M.\ 2016.\ Wormhole solutions sourced by fluids, II: three-fluid two-charged sources.\ European Physical Journal C 76. doi:10.1140/epjc/s10052-015-3836-4





\end{thebibliography}
\end{document}